\documentclass[aip,jmp,showpacs,twocolumn,nofootinbib]{revtex4}
\usepackage{graphicx}
\usepackage{dcolumn}
\usepackage{bm}
\usepackage{color}
\def\beq{\begin{eqnarray}}
\def\eeq{\end{eqnarray}}

\begin{document}

\title{Spectroscopy of annular drums and quantum rings: perturbative and nonperturbative results}
\author{Carlos Alvarado}
\affiliation{Facultad de Ciencias, Universidad de Colima, \\
Bernal D\'{\i}az del Castillo 340, Colima, Colima, Mexico} 
\author{Paolo Amore}
\email{paolo.amore@gmail.com (corresponding_author)}
\affiliation{Facultad de Ciencias, CUICBAS, Universidad de Colima, \\
Bernal D\'{\i}az del Castillo 340, Colima, Colima, Mexico} 

\begin{abstract}
We obtain systematic approximations to the states (energies and wave functions) of quantum rings (annular 
drums) of arbitrary shape by conformally mapping the annular domain to a simply connected domain. Extending
the general results of Ref.~\cite{Amore09} we obtain an analytical formula for the spectrum of quantum ring of
arbirtrary shape: for the cases of a circular annulus and of a Robnik ring considered here this formula is
remarkably simple and precise. We also obtain precise variational bounds for the ground state of 
different quantum rings.  Finally we extend the Conformal Collocation Method of \cite{Amore08,Amore09} to the
class of problems considered here and calculate precise numerical solutions for a large number of states ($\approx 2000$).
\end{abstract}
\maketitle

\section{Introduction}
\label{intro}

This paper extends the general results obtained in a recent paper by one of us, ref.~\cite{Amore09}, 
for simply connected drums and quantum billiards  to domains with a hole, i.e. annular drums or quantum rings.
Quantum rings are two dimensional regions of annular shape where an electron is confined and possibly subject to
an external field (magnetic or electric); as for the simply connected case studied in ref.~\cite{Amore09}, only few 
special cases may be solved exactly, such as for the circular annulus, where the solutions may be expressed in terms
of Bessel functions of first and second kind. However, quantum rings of general shape, for which exact solutions 
are not known, present interesting physical behaviors, thus justifying the effort of finding 
analytical or numerical approximations. 

For instance, it is known that the bendings of an infinite wire cause the appearance of bound states 
below the continuum threshold~\cite{Goldstone92} and of localized states in quantum rings in correspondence
of regions of maximum curvature~\cite{Gridin04}. In particular, Gridin and collaborators
have formulated in Ref.~\cite{Gridin04} an asymptotic approximation to the modes of a quantum ring, based 
on the assumption that the ratio of the ring half-width to the radius of curvature is small. Their approach
extends to ring like domain the classical Keller-Rubinow method~\cite{Keller60,Keller85}

In this paper we adopt a completely different strategy, which is both systematic and simple, and which can be used
to obtain precise analytical and/or numerical approximations for the energies and eigenfunctions of a given quantum ring.
The approach that we propose is based on a  generalization of the methods described in Ref.~\cite{Amore09} and 
allows one to solve Helmholtz equation for the quantum ring of arbitrary shape by conformally mapping the ring  to 
a simply connected domain. The results obtained in this way are very precise and prove to be useful even 
for the exactly solvable circular ring: in this case we have found an extremely precise formula for the energies 
of the rings, which avoids the use of Bessel functions and their zeroes. 

It is important to underline that the method that we propose is systematic and that it can be applied to rings 
of arbitrary width and curvature. 

The paper is organized as follows: in Section \ref{sec2} we describe the numerical implementation of our method, 
using the Conformal Collocation Method of \cite{Amore08,Amore09}; in Section \ref{sec3} we discuss the extension 
of the analytical techniques of \cite{Amore09} to the case of annular domains; in Section \ref{sec4} we
apply the methods, both analytical and numerical, to the solvable problem of a circular annulus and to the
less tractable problem of a "Robnik's ring", i.e. a ring whose external border corresponds to the family of 
quantum billiards studied by Robnik in \cite{Robnik84} and known as "Robnik's billiards". Finally in Section \ref{concl}
we summarize our results and draw our conclusions.

\section{Conformal collocation method}
\label{sec2}

In this section we describe the application of the Conformal Collocation Method (CCM) of Ref.~\cite{Amore08,Amore09} 
to the solution of Schr\"odinger equation in a quantum ring. Although this approach has been already 
described in detail in those papers, we briefly review it here to make the discussion self-contained
and to highlight the modifications which are needed to implement the specific problem at hand.

Our starting point is the homogeneous Helmholtz equation on a "ring-like" domain $\mathcal{D}$, which 
can be conformally mapped to an inhomogeneous Helmholtz equation on a "simpler" domain $\Omega$, which is 
assumed here to be a rectangle. Let $w = f(z)$ where $w = u + i v$ and $z = x + i y$ ($(u,v) \in \mathcal{D}$ 
and $(x,y) \in \Omega$).

As a result one is left to work with the inhomogeneous Helmholtz equation on $\Omega$:
\beq
- \frac{1}{\Sigma(x,y)} \Delta \psi(x,y) = E \psi(x,y) \ .
\label{B1}
\eeq
where $\Sigma \equiv \left| \frac{df}{dz} \right|^2$. 

\begin{figure}
\begin{center}
\bigskip\bigskip\bigskip
\includegraphics[width=6cm]{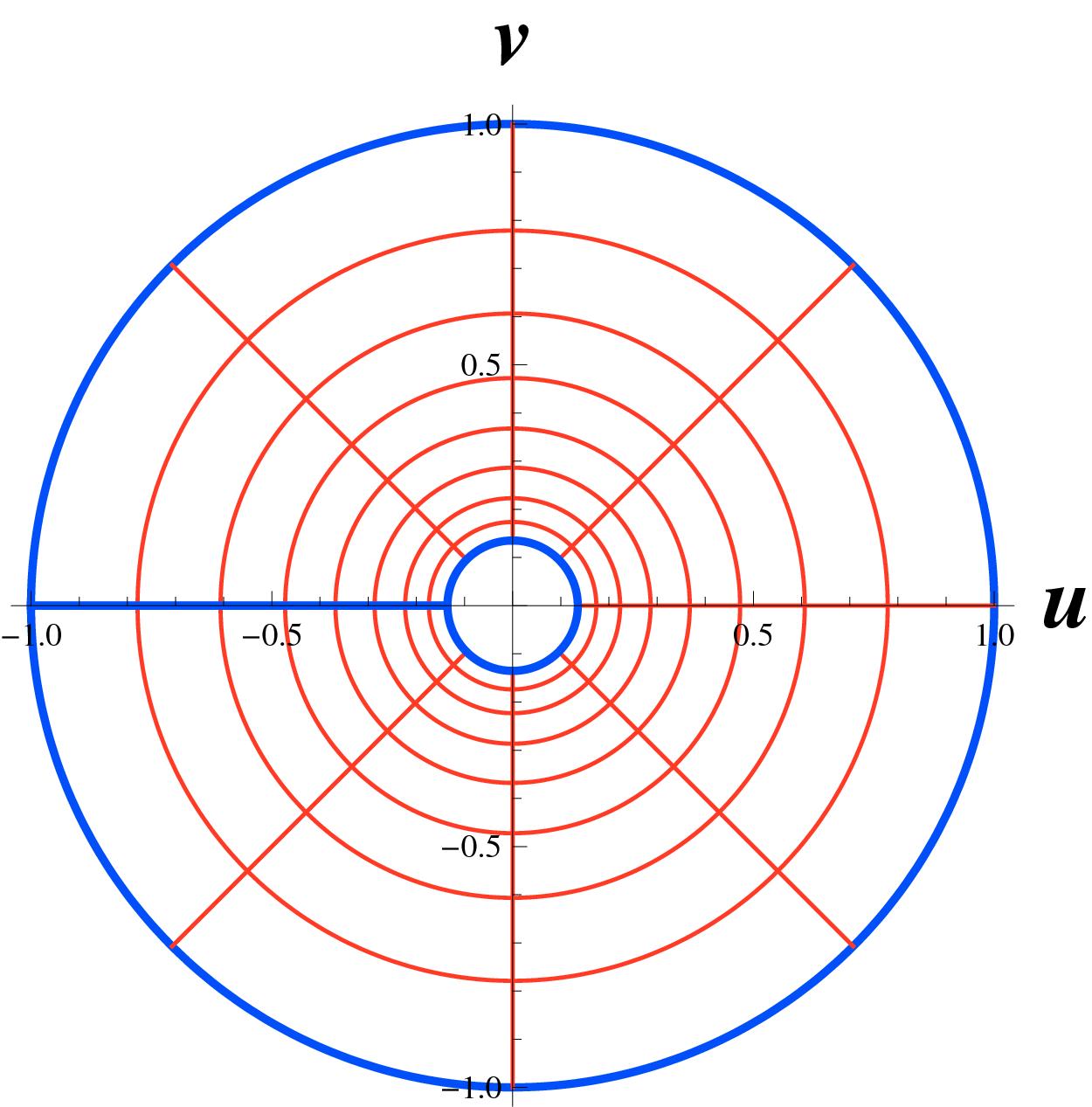}
\caption{(color online) Annulus obtained conformally mapping a rectangle of sides $2L_x=2$ and $2L_y=2 \pi$ 
centered in the origin.}
\label{Fig_1}
\end{center}
\end{figure}

An explicit example of conformal map with the desired properties is clearly the 
exponential map
\beq
f(z) = e^{z-L_x} \ ,
\label{cmap}
\eeq
which maps a rectangle of sides $2 L_x$ and $2 \pi$ centered in the origin into an annulus of 
radiuses $e^{-2 L_x}$ and $1$ respectively. Fig.~\ref{Fig_1} displays the annulus obtained 
mapping a rectangle of sides $L_x=2$ and $L_y=2 \pi$ centered in the origin (clearly 
using a smaller $L_y$ one would obtain an arc instead of the full ring). Notice that
the function of eq.~(\ref{cmap}) maps the horizontal sides into the horizontal segment 
which cuts the ring on the negative axis. This is clearly a different situation from those
considered in Ref.~\cite{Amore09}: as a matter of fact in this case the direct approach of 
Ref.~\cite{Amore09} would describe an annulus with a straight cut on which Dirichlet
boundary conditions are obeyed. Although this is also an interesting problem in itself, we 
want here to treat a ring with no cuts.

The solution to this problem is obtained by imposing periodic (Dirichlet) boundary conditions on the
horizontal (vertical) sides of the rectangle. Under the conformal map of eq.~(\ref{cmap}) the rectangle is then
mapped into an annulus fulfilling Dirichlet boundary conditions on the smaller and larger circles.
To implement the new boundary conditions in the problem we thus need to introduce a proper set of 
Little Sinc Functions (LSF): this set corresponds to $LSF_1$ of  Ref.~\cite{Amore09c}.

We report here the explicit form of the $LSF_1$:
\beq
s^{(I)}_k(N,L,x) &=& \frac{(-1)^k}{(N+1)} \frac{\sin\left(\frac{(N+1)\pi x}{2L}\right)}
{\sin\left(\frac{\pi x}{2L} - \frac{\pi k}{N+1}\right)}
\eeq
with $k=-N/2, \dots , N/2$. Notice that we use here a different convention for $N$ with respect 
to Ref.~\cite{Amore09c}. These functions define the homogeneous grid
$\displaystyle x_k = \frac{2Lk}{N+1}$ and satisfy the orthogonality relation
\beq
\int_{-L}^{+L} s^{(I)}_k(N,L,x) s^{(I)}_j(N,L,x) dx = \frac{2L}{N+1} \delta_{kj}.
\eeq
Notice that $h^{(I)} \equiv \frac{2L}{N+1}$ is the grid spacing of the $LSF_1$.

The $LSF_2$ fulfilling Dirichlet boundary conditions may be cast in the form
\beq
s^{(II)}_k(N,L,x) = \frac{(-1)^k}{N}
\frac{\cos\left(\frac{\pi k}{N}\right) \sin\left(\frac{N\pi x}{2L}\right)}
{\sin \left(\frac{\pi x}{2L}\right) - \sin \left(\frac{\pi k}{N}\right)}.
\eeq
with $k=-N/2+1, \dots, N/2-1$~\footnote{The reader may check that the expression given here 
for these function is equivalent to the one used before in Ref.~\cite{Amore09}}.
In this case these function define the homogeneous 
grid $\displaystyle x_k = \frac{2Lk}{N}$ and obey the orthogonality relation
\beq
\int_{-L}^{+L} s^{(II)}_k(N,L,x) s^{(II)}_j(N,L,x) dx = \frac{2L}{N} \delta_{kj}.
\eeq
Notice that $h^{(II)} \equiv \frac{2L}{N}$ is the grid spacing of the $LSF_2$.

For a given $N$ (even integer) there are $N+1$ ($N-1$) LSF functions obeying periodic (Dirichlet)
boundary conditions, each peaked (with value 1) at a point $x_k$ and vanishing at the remaining 
grid points $x_j$, $j\neq k$. 

A function $f(x)$ obeying periodic bc may be interpolated using the $s_k^{(I)}(h,N,x)$ as
\beq
f(x)  \approx \sum_{k=-N/2}^{N/2} f(x_k) s^{(I)}_k(h,N,x) \ .
\label{B2}
\eeq
Similarly we may derive twice this expression to obtain
\beq
\frac{d^2f(x)}{dx^2}  &\approx& \sum_{k=-N/2}^{N/2} f(x_k) \ \frac{d^2s^{(I)}_k(x)}{dx^2} \nonumber \\
 &\approx& \sum_{k=-N/2}^{N/2} \sum_{j=-N/2}^{N/2} f(x_k) \ \left. \frac{d^2s^{(I)}_k(x)}{dx^2}\right|_{x_j} 
s^{(I)}_j(h,N,x) \nonumber \\
&\equiv&  \sum_{k=-N/2}^{N/2} \sum_{j=-N/2}^{N/2} f(x_k) \ c_{kj}^{(I2)} \ 
s^{(I)}_j(h,N,x)  , 
\label{B3}
\eeq
where in the last line we have introduced the matrix $c_{kj}^{(I2)} \equiv \left. \frac{d^2s^{(I)}_k(x)}{dx^2}\right|_{x_j}$,
which provides a representation for the second derivative operator on the grid. The case of Dirichlet 
bc has already been discussed in Ref.~\cite{Amore09}, although it can be obtained straightforwardly 
repeating the same steps done here. To take into account the presence of different sets of LSF we modify 
the notation of Ref.~\cite{Amore09} and call $c_{kj}^{(II2)}$ the matrix elements of the second derivative
obtained with Dirichlet bc. Notice that the latin indices $k,j$ span different values in the two cases and 
that the grid points also differ in the two cases.

Omitting some trivial steps (see Ref.~\cite{Amore09} for more detail) we may now may easily discretize  
eqn.~(\ref{B1}) using Dirichlet bc in the $x$ direction and periodic bc in the $y$ direction. 
A suitable "basis" for this discretization is obtained with the direct product of the LSF in each direction:
\beq
&-& \frac{1}{\Sigma(x,y)} \Delta  s^{(II)}_k(h,N,x) s^{(I)}_{k'}(h,N,y) = - \sum_{jj'} 
\frac{1}{\Sigma(x_j,y_{j'})} \nonumber \\
&\times& \left[ c_{kj}^{(II2)} \delta_{k'j'} + \delta_{kj} c_{k'j'}^{(I2)} \right]  s^{(II)}_j(h,N,x) s^{(I)}_{j'}(h,N,y) 
\eeq
where it is understood that the latin indices span different ranges for the two LSF.

To obtain the matrix element of the operator on the grid we need to associate a single integer to any pair of
indices which define an element of the two dimensional grid. 
We may write:
\beq
k &=& K-(N-1)\left[ \frac{K}{N-1+\epsilon} \right]-\frac{N}{2} \\
k'&=& -\frac{N}{2}+\left[ \frac{K}{N-1+\epsilon} \right]  \ ,
\eeq
where $\epsilon \rightarrow 0^+$ and $\left[ a \right]$ means integer part of $a$. In this way we are
able to identify a point of the grid in terms of a single integer $K$, which takes values from $1$ to 
$N^2-1$.

Using these relations we may read off the matrix element of $\hat{O}$ as
\beq
O_{KK'} = -  \frac{1}{\Sigma(x_j,y_{j'})}  \left[ c_{kj}^{(II2)} \delta_{k'j'} + \delta_{kj} c_{k'j'}^{(I2)} \right]  \ .
\eeq

Although the procedure described above uses grid with the same $N$ for the $x$ and $y$ direction, a more appropriate 
choice  is to use meshes with the same grid size on each orthogonal direction, i.e. $h^{(I)} \approx h^{(II)}$.
In this way we may establish the relation between the number of grid points on each direction
\beq
N_x  \approx  \frac{L_x}{L_y} \ (N_y+1) \nonumber \ .
\eeq

We may easily understand the physics contained in this relation: for thin rings, $L_x \ll L_y$, the number of grid points 
in the $x$-direction is much less than the number of grid points in the $y$-direction, since the excitation of trasverse 
modes requires much higher energy than the excitation of longitudinal modes. 

In this case one can extend the previous relations for the grid to:
\beq
k &=& K-(N_{x}-1)\left[ \frac{K}{N_{x}-1+\epsilon} \right]-\frac{N_{x}}{2} \nonumber \\
k'&=& -\frac{N_{y}}{2}+\left[ \frac{K}{N_{x}-1+\epsilon} \right] \nonumber \ .
\eeq

The implementation of these considerations
allows us to represent the differential operator as a $(N_x-1) (N_y+1) \times (N_x-1) (N_y+1)$ hermitean matrix.

It is useful to summarize the differences with the results of Ref.~\cite{Amore09}:
\begin{itemize}
\item The region $\Omega$ is a rectangle of sides $2L_x$ and $2L_y$, where $L_y=\pi$ to allow a closed
ring

\item The operator $\hat{O}$ is represented on the grid by a $(N_x-1) (N_y+1) \times (N_x-1) (N_y+1)$ hermitean matrix

\item The collocation points, corresponding to the nodes of the LSF functions, differ for the $x$ and $y$
directions: on the $x$ direction they are distributed following the zeroes of the LSF with Dirichlet boundary 
conditions, while on the $y$ direction they are distributed following the zeroes of the LSF with periodic boundary 
conditions. Their numbers ($N_x$ and $N_y$) are also different, although the grid size is (approximately) the same.

\end{itemize}

Apart from these small differences, the remaining features of the collocation method 
are unchanged. In particular, the matrix representing $\hat{O}$ on the grid is obtained
from the product of a diagonal matrix, representing $1/\Sigma$ on the grid, with a non-diagonal 
sparse matrix, representing the Laplacian with mixed bc on the grid. While the first matrix is specific 
to the problem considered and therefore it needs to be calculated each time that a different shape is
chosen, the second matrix is universal and therefore it can be calculated and stored once and for all.
As before the calculation of the diagonal matrix is not computationally demanding since it 
just requires the evaluation of the function $1/\Sigma$ at the $(N_x-1) (N_y+1)$ points forming the grid.

We will illustrate later specific applications of the CCM.

\section{Analytical methods}
\label{sec3}

In this section we generalize the analytical approach of Ref.~\cite{Amore09} to describe 
quantum rings. We may consider two different approaches: in the first one the quantum ring is 
obtained performing a conformal map of a rectangle centered in the origin (as described in the
previous section); in the second one the quantum ring is obtained by applying a conformal map 
directly to a circular annulus. 

While the formulas of Ref.~\cite{Amore09} hold both cases, one needs to work with different
orthonormal basis in the two cases.

In the first case where $\Omega$ is a rectangle of sides $2L_x$ and $2\pi$, the basis is 
obtained by the direct product of functions obeying Dirichlet and periodic boundary 
conditions, i.e.
\beq
\psi_{n_x}(x) = \frac{1}{\sqrt{L_x}} \ \sin \left(\frac{n_x \pi}{2L_x}(x+L_x)\right)
\eeq
for the Dirichlet bc and
\beq
\chi_0(y)   &=& \frac{1}{\sqrt{2\pi}}, \\ 
\chi_{n_y}(y) &=& \frac{1}{\sqrt{\pi}} \cos \left(n_y y\right), \\ 
\phi_{n_y}(y) &=& \frac{1}{\sqrt{\pi}} \sin \left(n_y y\right) ,
\eeq
for the periodic bc. Notice that $n_{x,y}=1,2,\dots$. 

The basis on $\Omega$ may then be written as
\beq
\Psi_{n_x,n_y,s}(x,y) = \psi_{n_x}(x) \times \left\{ \begin{array}{ccc} \chi_{ny}(y) & , & s=1 \nonumber \\ \phi_{ny}(y) &, & s=2\end{array} \right. \ ,
\eeq
where the value $n_y=0$ can only be reached for the states with $s=1$.

In the second case, the wave function of a circular annulus with $a<r<b$ is (see for example Ref.~\cite{Kuttler84})
\beq
&& \Phi_{m,n,s}(r,\theta) = N_{mns} \ \left[ Y_m\left(k_{mn}\right) J_m\left(\frac{k_{mn}r}{a}\right) \right. \nonumber \\
&-& \left.
J_m\left(k_{mn}\right) Y_m\left(\frac{k_{mn}r}{a}\right) \right]  \times \left\{ \begin{array}{ccc} \cos n \theta & , & s=1 \nonumber \\ \sin n\theta &, & s=2\end{array} \right. \ , 
\eeq
where $J_m$ and $Y_m$ are Bessel functions of first and second kind and $k$ is  
the $n^{th}$ root of the equation
\beq
Y_m\left(k\right) J_m\left(\frac{kb}{a}\right) - J_m\left(k\right) Y_m\left(\frac{k b}{a}\right) = 0 \ .
\eeq

The energies of the annulus are then given by
\beq
E_{mn} = \left(\frac{k_{mn}}{a}\right)^2 \ , \ m=0,1,2,\dots  \ , \ n= 1,2\dots
\label{exact_energies}
\eeq

Although one may work equally well with each of the two basis, from the point of view of an analytical calculation the first one offers 
the advantage of simplicity, since it involves only elementary functions. We will therefore focus on this basis.

We consider the most general conformal transformation which maps the rectangle $\Omega$ onto a ring of arbitrary shape:
\beq
g(z) = C \sum_{k=0}^\infty \eta_k \ \left(e^{z-L_x}\right)^{k+1} \equiv C \ \bar{g}(z) \ ,
\eeq
where $\eta_0=1$ and $C>0$ is a constant factor, representing a dilation. The energies of the ring obtained using 
the mapping $g(z)$ ($E_n$) are related to those of the ring obtained using $\bar{g}(z)$ ($\bar{E}_n$) by the simple
relation~\footnote{Here the index $n$ represents all the set of quantum numbers defining the state and not an individual
quantum number.}
\beq
E_n = \frac{\bar{E}_n}{C^2} \nonumber \ .
\eeq
We may therefore work with $\bar{g}(z)$ and then simple rescale the energies obtained.

Notice also that with the choice $C=1$ and $\eta_{k>0}=0$ we obtain the map to the circular annulus considered earlier.

With simple algebra we obtain the conformal density
\beq
\Sigma(x,y) &\equiv& \left| \frac{d\bar{g}}{dz}\right|^2 = \sum_{k=0}^\infty \sum_{j=0}^\infty \eta_k \eta_j (k+1) (j+1) \nonumber \\
&\cdot& \left(e^{x-L_x}\right)^{(k+j)+2} \ \cos \left(y (k-j)\right) 
\eeq
and $\sigma(x,y) = \Sigma(x,y)-1$.

As discussed in Ref.~\cite{Amore09} we need to calculate the matrix elements of $\sigma$ between the states of $\Omega$.
We find:
\beq
&& \langle n_x,n_y,s | \sigma(x,y) | n'_x,n'_y,s'\rangle = -\delta_{n_x,n'_x} \delta_{n_y,n'_y} \delta_{ss'} \nonumber \\
&+& \sum_{k=0}^\infty \sum_{j=0}^\infty \eta_k \eta_j (k+1) (j+1) \Delta_{n_x,n_y,s,n'_x,n'_y,s'}(k,j)
\eeq
\begin{widetext}
where 
\beq
 \Delta_{n_x,n_y,s,n'_x,n'_y,s'}(k,j) &\equiv& \int_{-L_x}^{+L_x} dx 
\int_{-\pi}^{+\pi} dy \ \Psi_{n_x,n_y,s}(x,y) \left(e^{x-L_x}\right)^{(k+j)+2} \ \cos \left(y (k-j)\right) \Psi_{n'_x,n'_y,s'}(x,y)  \nonumber \\
&=& \alpha_{n_y} \alpha_{n'_y} \delta_{ss'} W_{n_x,n'_x}(k,j) \mathcal{I}^{(s)}_{n_y,n'_y,k,j} \ ,
\eeq
\end{widetext}
and $\alpha_{0} = 1/\sqrt{2\pi} \delta_{s,1}$ and $\alpha_{n_y>0} = 1/\sqrt{\pi}$. 

\bigskip

\begin{widetext}
We have also introduced the definitions
\beq
W_{n_x,n'_x,k,j} &\equiv& \int_{-L_x}^{+L_x}  \psi_{n_x}(x) \left(e^{x-L_x}\right)^{(k+j)+2 } \psi_{n'_x}(x) dx \nonumber \\
&=& 
\frac{8 \pi ^2 L_x {n_x} {n'_x} (j+k+2) e^{-2 L_x (j+k+2)} \
\left((-1)^{{n_x}+{n'_x}} e^{2 L_x (j+k+2)}-1\right)}{\left(4 L_x^2 \
(j+k+2)^2+\pi ^2 ({n_x}-{n'_x})^2\right) \left(4 L_x^2 (j+k+2)^2+\pi \
^2 ({n_x}+{n'_x})^2\right)} 
\eeq
and
\beq
\mathcal{I}^{(1)}_{n_y,n'_y,k,j} &\equiv& 
\int_{-\pi}^{+\pi} \cos (n_y y) \cos (n'_y y) \ \cos ((k-j) y) dy \nonumber \\
&=&  \frac{\pi}{2} \left[ \delta_{n_y-n'_y+(k-j)} + 
\delta_{n_y+n'_y+(k-j)} + \delta_{n_y-n'_y-(k-j)} +\delta_{n_y+n'_y-(k-j)} \right]\nonumber \\
\mathcal{I}^{(2)}_{n_y,n'_y,k,j} &\equiv& \int_{-\pi}^{+\pi} \sin (n_y y) \sin (n'_y y) \ \cos ((k-j) y) dy \nonumber \\
&=&  \frac{\pi}{2} \left[ \delta_{n_y-n'_y+(k-j)} - 
\delta_{n_y+n'_y+(k-j)} + \delta_{n_y-n'_y-(k-j)} -\delta_{n_y+n'_y-(k-j)} \right] \ . \nonumber \\
\eeq  
\end{widetext}
Notice that the $\mathcal{I}^{(s)}_{n_y,n'_y,k,j}$  have been already defined in Ref.~\cite{Amore09}; the reader should also 
observe that there are no terms mixing states with different values of $s$ and that $\mathcal{I}^{(1)}_{n_y,n'_y,k,j} = 
\mathcal{I}^{(2)}_{n_y,n'_y,k,j}$ unless $n_y+n'_y = \pm (k-j)$.

We may understand the physical consequences of these properties by considering the approximate expression for the energy:
\beq
\bar{E}_{n_x,n_y,s} \approx \frac{\epsilon_{n_x,n_y,s}}{\langle n_x,n_y,s | \Sigma(x,y) | n_x,n_y,s\rangle} \ ,
\label{weyl}
\eeq
where
\beq
\epsilon_{n_x,n_y,s} \equiv  \frac{n_x^2\pi^2}{4L_x^2}+ n_y^2
\eeq
are the "unperturbed" energies (i.e. the energy on the rectangle $\Omega$). 

This expression has been derived in Ref.~\cite{Amore09} resumming specific terms in the perturbative expansion 
corresponding to a geometric series: it has later been applied in Ref.~\cite{Amore09b} to obtain a Weyl-like law for inhomogeneous drums.

\bigskip

\begin{widetext}
With simple algebra we obtain
\beq
\langle n_x,n_y,s | \Sigma(x,y) | n_x,n_y,s\rangle &=& \sum_{k=0}^\infty \sum_{j=0}^\infty \eta_k \eta_j (k+1) (j+1)  \ \left[
W_{n_x,n_x,k,k} \delta_{kj} + (-1)^{s+1} W_{n_x,n_x,k,j} \delta_{2 n_y-|k-j|}\right] \nonumber \\
&=& \sum_{k=0}^\infty \eta_k^2 (k+1)^2 W_{n_x,n_x,k,k} + (-1)^{s+1} \sum_{k=2 n_y}^\infty \eta_k \eta_{k-2n_y} (k+1) (k+1-2 n_y) 
W_{n_x,n_x,k,k-2n_y} \nonumber \\
&+&  (-1)^{s+1} \sum_{k=0}^\infty \eta_k \eta_{k+2n_y} (k+1) (k+1+2 n_y)  W_{n_x,n_x,k,k+2n_y} \ ,
\eeq
\end{widetext}
where the terms depending explicitly on $s$ will be responsible of the breaking of the degeneration
of levels.

\section{Applications}
\label{sec4}

We will now consider few applications of the numerical and analytical techniques developed here and 
in Ref.~\cite{Amore09} to quantum rings of different shape.

\subsection{Circular annulus}

Our first application is to the calculation of the energies and wave functions of the circular annulus using the
basis of the rectangle.  

We may use eq.~(\ref{weyl}) and obtain an explicit formula for the energies of a circular annulus:
\beq
E_{n_x,n_y,s} \approx  \frac{2 \left(\log^2(a)+\pi ^2 {n_x}^2\right) \left({n_y}^2 \log^2(a)+
\pi^2 {n_x}^2\right)}{\pi ^2 \left(a^2-1\right) {n_x}^2 \log (a)}  ,
\label{en_annulus_weyl}
\eeq
where $a = e^{-2L_x}$ is the inner radius of the ring. 

Notice that the states corresponding to $n_y=0$ are non-degenerate,
whereas the states corresponding to $n_y>0$ are doubly degenerate: in other words the basis that we are using reproduces the
exact pattern of degeneration of the circular annulus.

\begin{figure}[hbt]
\begin{center}
\bigskip\bigskip\bigskip
\includegraphics[width=8cm]{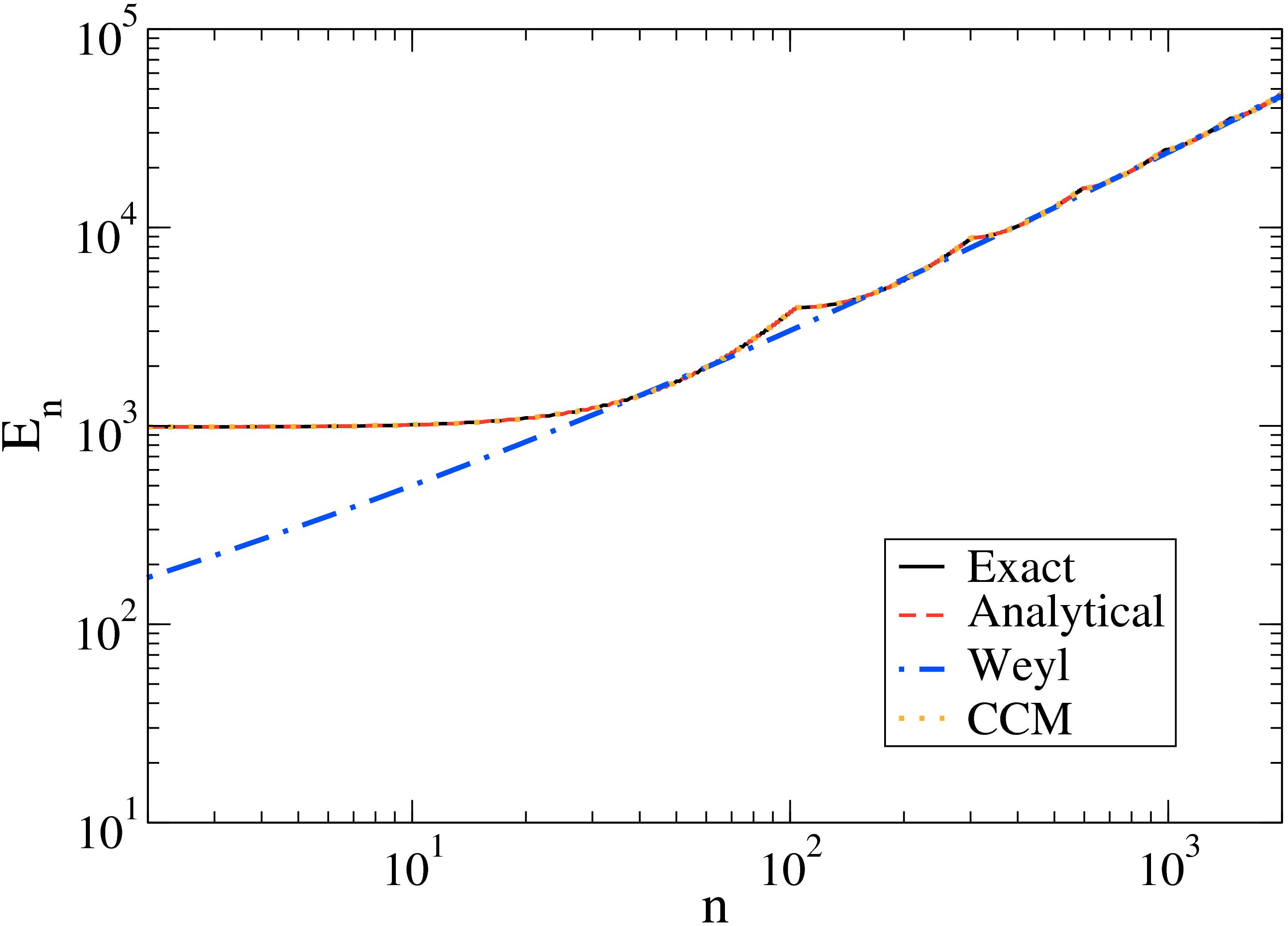}
\caption{(color online) Energy of the first 2000 states of an annulus with $a=9/10$ and $b=1$.
The solid line are the exact results of  eq.~(\ref{exact_energies}); the dashed line corresponds to
eq.~(\ref{en_annulus_weyl}); the dotted line are the numerical results obtained with CCM; the dot-dashed line
are the results obtained with Weyl's law supplemented by Weyl's conjecture.}
\label{Fig_3}
\end{center}
\end{figure}

In Fig.~\ref{Fig_3} we compare the exact numerical results obtained solving  eq.~(\ref{exact_energies}) (solid curve) with the 
results obtained using the analytical formula (\ref{en_annulus_weyl}) (dashed curve) for the first $2000$ states. 
The dotted line corresponds to the numerical results obtained using the CCM with a grid with $N_x=14$ and $N_y=400$ (remember 
that the ratio $N_x/(N_y+1)$ should approximately be $L_x/L_y$).
Finally the dot-dashed line are the results obtained using Weyls'law supplemented by Weyl's conjecture
\beq
E_n \approx \frac{4\pi n}{A}  + \frac{L}{A} \sqrt{\frac{4 \pi n}{n}} + \dots \ ,
\label{weyls_conj}
\eeq
where $A$ and $L$ are area and perimeter of the ring respectively (Dirichlet boundary conditions are chosen).

\begin{figure}[hbt]
\begin{center}
\bigskip\bigskip\bigskip
\includegraphics[width=8cm]{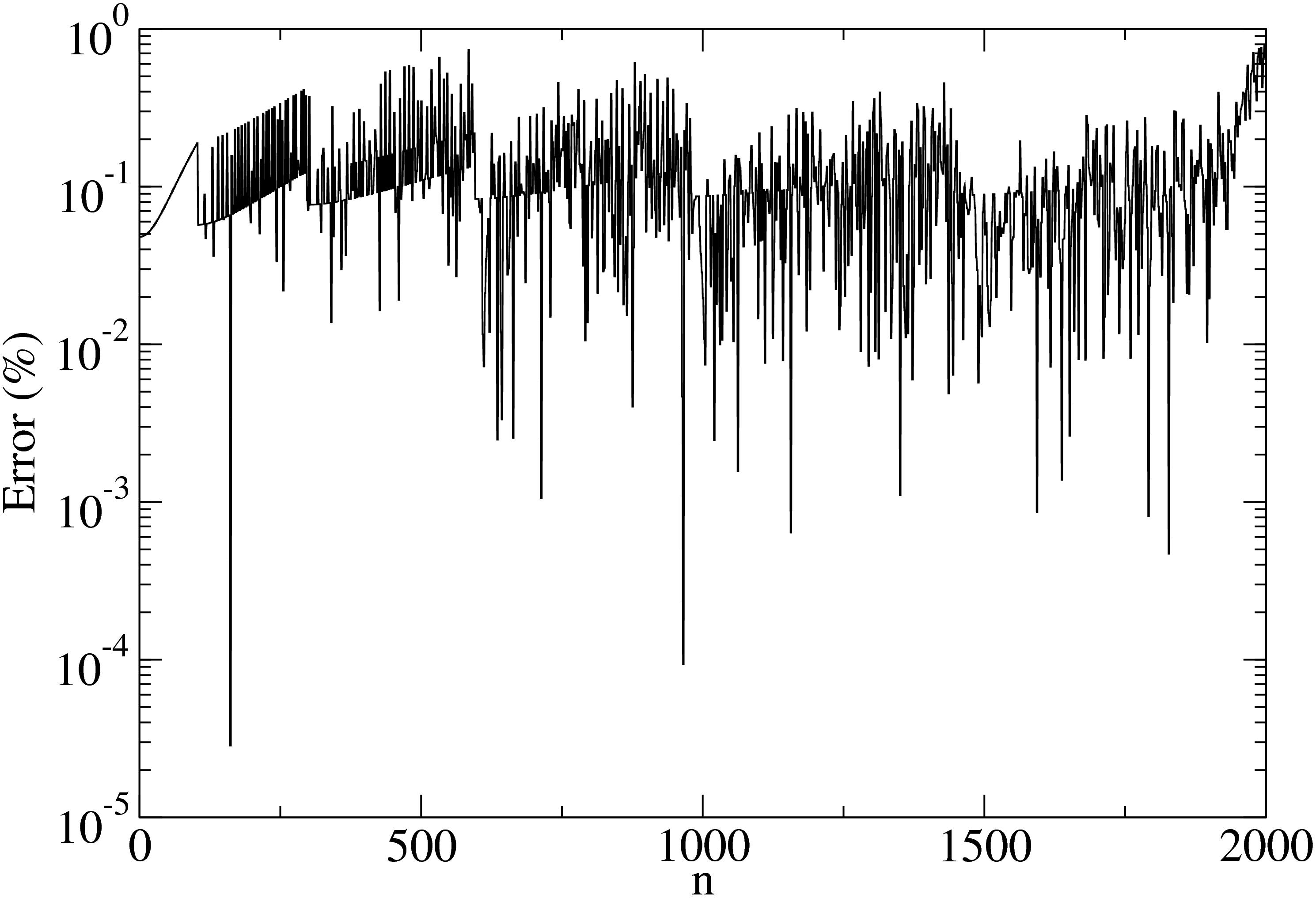}
\caption{(color online) Relative error of the analytical formula (\ref{en_annulus_weyl})
with respect to the exact results}
\label{Fig_4}
\end{center}
\end{figure}

Amazingly the first three curves are practically indistinguishable.
In Fig.~\ref{Fig_4} we display the the error of the analytical formula,  which is about $0.1 \ \%$ for
all the states considered.

We may also obtain precise results for the ground state of the ring using a variational approach. Following Ref.~\cite{Amore09}
we pick a trial state
\beq
|\chi \rangle = \sum_k c_k^{(0)} | k\rangle \ ,
\eeq
where $|k\rangle \equiv | k_x,k_y\rangle$ are eigenstates of the rectangle with mixed bc. Using the inverse 
operator $\hat{O}^{-1}$ we may generate a new state
\beq
|\Psi_0^{(1)} \rangle = \hat{O}^{-1} |\chi\rangle \ ,
\eeq
which has a larger overlap with the true ground state of the system. 

The expectation value of $\hat{O}$ in this state is 
\beq
E_0^{(1)} &=& \frac{\langle \Psi_0^{(1)} | \hat{O} | \Psi_0^{(1)}\rangle}{\langle \Psi_0^{(1)} | 
\Psi_0^{(1)}\rangle} \ ,
\label{C8}
\eeq
where
\beq
\langle \Psi_0^{(1)} | \hat{O} | \Psi_0^{(1)}\rangle &=& 
\sum_{k,l,m=0}^\infty  \frac{c_k^{(0)} c_m^{(0)}}{\epsilon_l } \langle m | \Sigma^{1/2} | l \rangle  
\langle l | \Sigma^{1/2} | k \rangle \\
\langle \Psi_0^{(1)} | \Psi_0^{(1)}\rangle &=& 
\sum_{l,l',m,m'=0}^\infty  \frac{c_m^{(0)} c_{m'}^{(0)}}{\epsilon_l \epsilon_{l'} } 
\langle m' | \Sigma^{1/2} | l' \rangle \langle l' | \Sigma | l \rangle  \nonumber \\
&\times& \langle l | \Sigma^{1/2} | m \rangle \ .
\label{C9}
\eeq

By minimizing $E_0^{(1)}$ with respect to the coefficients $c_k^{(0)}$ one can now find precise estimates for the
ground state energy. Let us now pick a specific trial state. We will work in the limit of a annulus with
$(b-a) \ll 1$, which corresponds to using $L_x \ll 1$. 

We use the trial state
\beq
|\chi \rangle &=& c_{0}^{(0)}  \psi_1(x) \chi_0(y) \nonumber \\
&+& \psi_1(x) \sum_{n_y=1}^{N} 
\left[ c_{2 n_y-1}^{(0)}  \phi_{n_y}(y) + c_{2 n_y}^{(0)}  \chi_{n_y}(y)\right]
\eeq
neglecting in first approximation the excited states in the $x$ direction.

We need to calculate the matrix elements
\beq
\langle m | \Sigma^{1/2} | l \rangle  &=&  \langle 1,m_y | \Sigma^{1/2} | 1, l_y \rangle  \nonumber \\
&=& \left[\int_{-L_x}^{+L_x} \psi_1^2(x) e^{x-L_x} dx \right] \delta_{m_y l_y} \nonumber \\
&=& \frac{\pi ^2 e^{-L_x} \sinh (L_x)}{L_x^2+\pi ^2} \delta_{m_y l_y} \\
\langle m | \Sigma | l \rangle  &=&  \langle 1,m_y | \Sigma | 1, l_y \rangle  \nonumber \\
&=& \left[\int_{-L_x}^{+L_x} \psi_1^2(x) e^{2 x-2 L_x} dx \right] \delta_{m_y l_y} \nonumber \\
&=& \frac{\pi ^2 \left(1-e^{-4 L_x}\right)}{4 \left(4 L_x^2+\pi ^2\right)} \delta_{m_y l_y}
\eeq

Therefore
\beq
E_0^{(1)} &=& \frac{\sum_k \frac{\left(c_k^{(0)}\right)^2}{\epsilon_k} 
\left[\frac{\pi ^2 e^{-L_x} \sinh (L_x)}{L_x^2+\pi ^2}\right]^2}{\sum_k \frac{\left(c_k^{(0)}\right)^2}{\epsilon_k^2} 
\left[\frac{\pi ^2 e^{-L_x} \sinh (L_x)}{L_x^2+\pi ^2}\right]^2 \frac{\pi ^2 \left(1-e^{-4 L_x}\right)}{4 \left(4 L_x^2+\pi ^2\right)} } \nonumber \\
&=& \frac{1}{\frac{\pi ^2 \left(1-e^{-4 L_x}\right)}{4 \left(4 L_x^2+\pi ^2\right)} }
\frac{\sum_k \frac{\left(c_k^{(0)}\right)^2}{\epsilon_k} }{\sum_k \frac{\left(c_k^{(0)}\right)^2}{\epsilon_k^2}}
\label{variational}
\eeq
where
\beq
\epsilon_k = \frac{\pi^2}{4 L_x^2} + \left[\frac{k+1}{2}\right]^2
\eeq
where $\left[\frac{k}{2}\right]$ is the integer part of $k/2$. Since one of the coefficients $c_k^{(0)}$ can
be chosen arbitrarily we may pick $c_0^{(0)} =1$ and minimize $E_0^{(1)}$ with respect to the remaining
coefficients. The result is $c_k^{(0)} = 0$ , for $k\geq 1$, which is equivalent to working with eq.~(\ref{weyl}).
This fact should not be surprising, since the states corresponding to non-vanishing $c_k^{(0)}$ (for $k\geq 1$)
break the rotational invariance of the true ground state of the annulus and therefore they must raise the energy.
This explains the very good precision of eq.~(\ref{en_annulus_weyl}) for annuli with $b-a \gg 1$.

To improve this formula for states with smaller inner radius we may form a variational trial state with 
a superposition of wave functions in the $x$ direction:
\beq
|\chi \rangle &=&\phi_0(y) \sum_{n_x=1}^{N} c_{n_x}^{(0)}  \psi_{n_x}(x)
\label{expansion}
\eeq

\begin{widetext}
Therefore we may write the matrix elements:
\beq
\langle m | \Sigma^{1/2} | l \rangle  &=&  \langle m_x,0 | \Sigma^{1/2} | l_x,0 \rangle  
= \left[\int_{-L_x}^{+L_x} \psi_{m_x}(x) \psi_{l_x}(x) e^{x-L_x} dx \right] \nonumber \\
&=& \left\{ 
\begin{array}{ccc} \frac{\pi ^2 e^{-L_x} m_x^2 \sinh (L_x)}{L_x^3+\pi ^2 L_x m_x^2} & , & m_x = l_x \\
\frac{8 \pi ^2 e^{-2 L_x} L_x m_x l_x \left(e^{2 L_x}-(-1)^{m_x+l_x}\right) \cos (\pi  (m_x+l_x))}{16 L_x^4+8 
\pi^2 L_x^2 \left(m_x^2+l_x^2\right)+\pi^4 \left(m_x^2-l_x^2\right)^2} & , & m_x \neq l_x \nonumber \\ 
\end{array} \right. \nonumber \\
\langle m | \Sigma | l \rangle  &=&  \langle m_x,0 | \Sigma | l_x,0 \rangle  
= \left[\int_{-L_x}^{+L_x} \psi_{m_x}(x) \psi_{l_x}(x) e^{2 x-2 L_x} dx \right] \nonumber \\
&=& \left\{ 
\begin{array}{ccc} 
\frac{\pi ^2 e^{-4 L_x} \left(e^{4 L_x}-1\right) m_x^2}{4 \left(4 L_x^3+\pi ^2 L_x m_x^2\right)}
& , & m_x = l_x \\
\frac{16 \pi ^2 e^{-4 L_x} L_x {l_x} {m_x} \left(e^{4 L_x}-(-1)^{{l_x}+{m_x}}\right) \cos (\pi({l_x}+{m_x}))}{256 L_x^4+32 
\pi ^2 L_x^2 \left({l_x}^2+{m_x}^2\right)+\pi ^4 \left({m_x}^2-{l_x}^2\right)^2} & , & m_x \neq l_x \nonumber \\ 
\end{array} \right. \nonumber \\
\eeq
Notice that in this case $\epsilon_n = \frac{\pi^2 n_x^2}{4 L_x^2}$.
\end{widetext}

In Table \ref{table-1} we report the ground state energy of the annulus calculated using the
variational principle, i.e  using the variational formula (\ref{variational}) together with 
eq.~(\ref{expansion}), with different numbers of variational parameters. The case $N=1$ corresponds 
to the simple analytical formula (\ref{en_annulus_weyl}), which is seen to work very well for larger
values of $a$. It is easy to understand why more and more terms are needed as $a$ gets smaller
and smaller: since $L$ gets larger, the conformal map strongly deforms the radial
coordinate in the annulus. Looking at Fig.~\ref{Fig_1}, which corresponds to $a=1/e^2$, we observe 
that the radial grid spacing is finer for smaller values of $r$. Therefore one needs more terms
in eq.~(\ref{variational}) to obtain good estimates of the energy. Notice that this argument
also holds for the Conformal Collocation Method, since also the collocation points are also 
distributed non-uniformly in the radial direction, as the central hole is made smaller.

\bigskip

\begin{widetext}
\begin{center}
\begin{table}[!htb]
\begin{tabular}{|c||c|c|c|c|c|c||c|}
	\hline
$a$	  &  $N=1$      &   $N=2$     & $N=3$       & $N=4$       & $N=5$       & $N=10$      & exact      \\
\hline \hline
$1/100000$& 24.74037682 & 13.58247813 & 10.13874976 & 8.585413233 & 7.763566890 & 6.639748109 & 6.493374419 \\ 
$1/10000$ & 20.56383831 & 11.77986914 & 9.129597650 & 7.994645335 & 7.425906261 & 6.751169408 & 6.693191961 \\
$1/1000$  & 16.67307038 & 10.21973107 & 8.409275276 & 7.691356864 & 7.367206060 & 7.062604524 & 7.048038125 \\ 
$1/100$   & 13.49800490 & 9.359102734 & 8.360288280 & 8.036684452 & 7.920642510 & 7.846777753 & 7.845162680 \\
$1/10$    & 13.31090873 & 11.31798812 & 11.04543198 & 10.99627372 & 10.98600010 & 10.98222952 & 10.98218981 \\
$1/2$     & 39.81860401 & 39.03110135 & 39.01558313 & 39.01353082 & 39.01336954 & 39.01328910 & 39.01328850 \\
$9/10$    & 987.1575032 & 986.6840018 & 986.6837734 & 986.6831548 & 986.6831530 & 986.6831306 & 986.6831305 \\
$99/100$  & 98696.22334 & 98695.79206 & 98695.79206 & 98695.79151 & 98695.79151 & 98695.79149 & 98695.79149 \\
	\hline
\end{tabular}
\caption{Energy of the ground state of the annulus with varying inner radius calculated using
the variational formula (\ref{variational}) together with eq.~(\ref{expansion}), with
different numbers of variational parameters ($N-1$ is the number of variational parameters) . 
Notice that in this case $\epsilon_n = \frac{\pi^2 n_x^2}{4 L_x^2}$.}
\label{table-1}
\end{table}
\end{center}
\end{widetext}

\begin{figure}[hbt]
\begin{center}
\bigskip\bigskip\bigskip
\includegraphics[width=8cm]{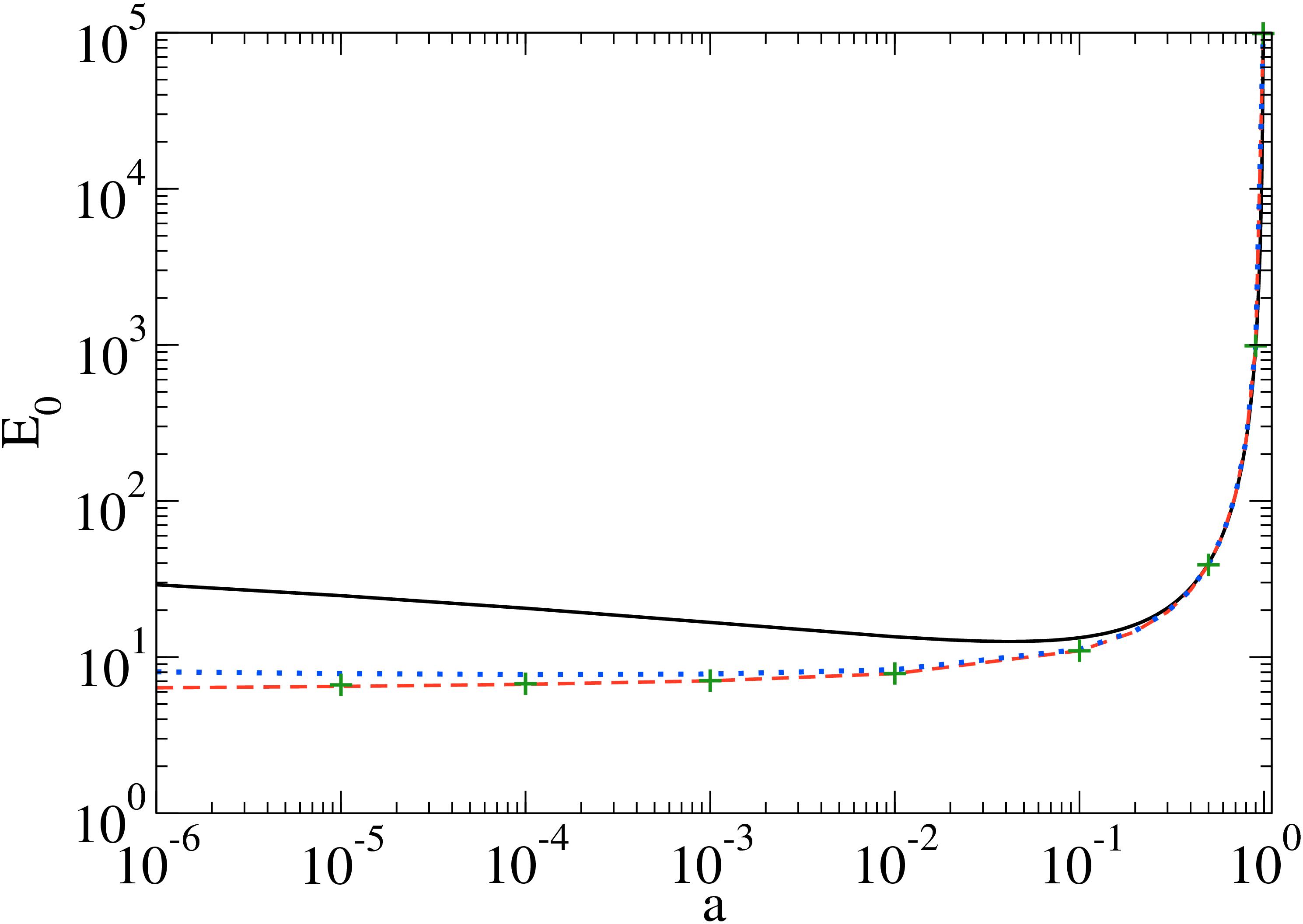}
\caption{(color online) Ground state energy of the annulus as a function of the inner radius keeping the outer radius 
fixed ($b=1$). The dashed line is the exact result, while the dashed line is the analytical formula of 
eq.~(\ref{en_annulus_weyl}). The dotted line is the result obtained using CCM with $N=50$; the pluses are the variational results
of Table \ref{table-1} corresponding to $N=10$.}
\label{Fig_2}
\end{center}
\end{figure}

In Fig.~\ref{Fig_2} we have plotted the exact energy of the ground state of the annulus as a function of the inner radius $a$
and we have compared it with the different approximations considered earlier. The simple analytical formula 
of eq.~(\ref{en_annulus_weyl}) is seen to work quite well for thin annuli, while the most accurate variational calculation
of Table \ref{table-1} is seen to reproduce the exact results even for very small inner radiuses. We also report the CCM results
corresponding to a grid with $N=50$.

\subsection{Robnik's rings}

We consider the conformal map
\beq
f(z) = e^{-L_x+z} + \alpha e^{-2 L_x+2 z}
\label{cardioid}
\eeq
which maps the rectangle of sides $2L_x$ and $2\pi$ into a deformed ring, which for $\alpha=0$ has the shape of a circular ring
and for $\alpha=1/2$ has the shape of a cardiod ring. We have called the family of these annular billiards "Robnik's rings",
in analogy with the simply connected billiards known as "Robnik's billiards" having the same external contour~\cite{Robnik84}.

\begin{figure}
\begin{center}
\bigskip\bigskip\bigskip
\includegraphics[width=6cm]{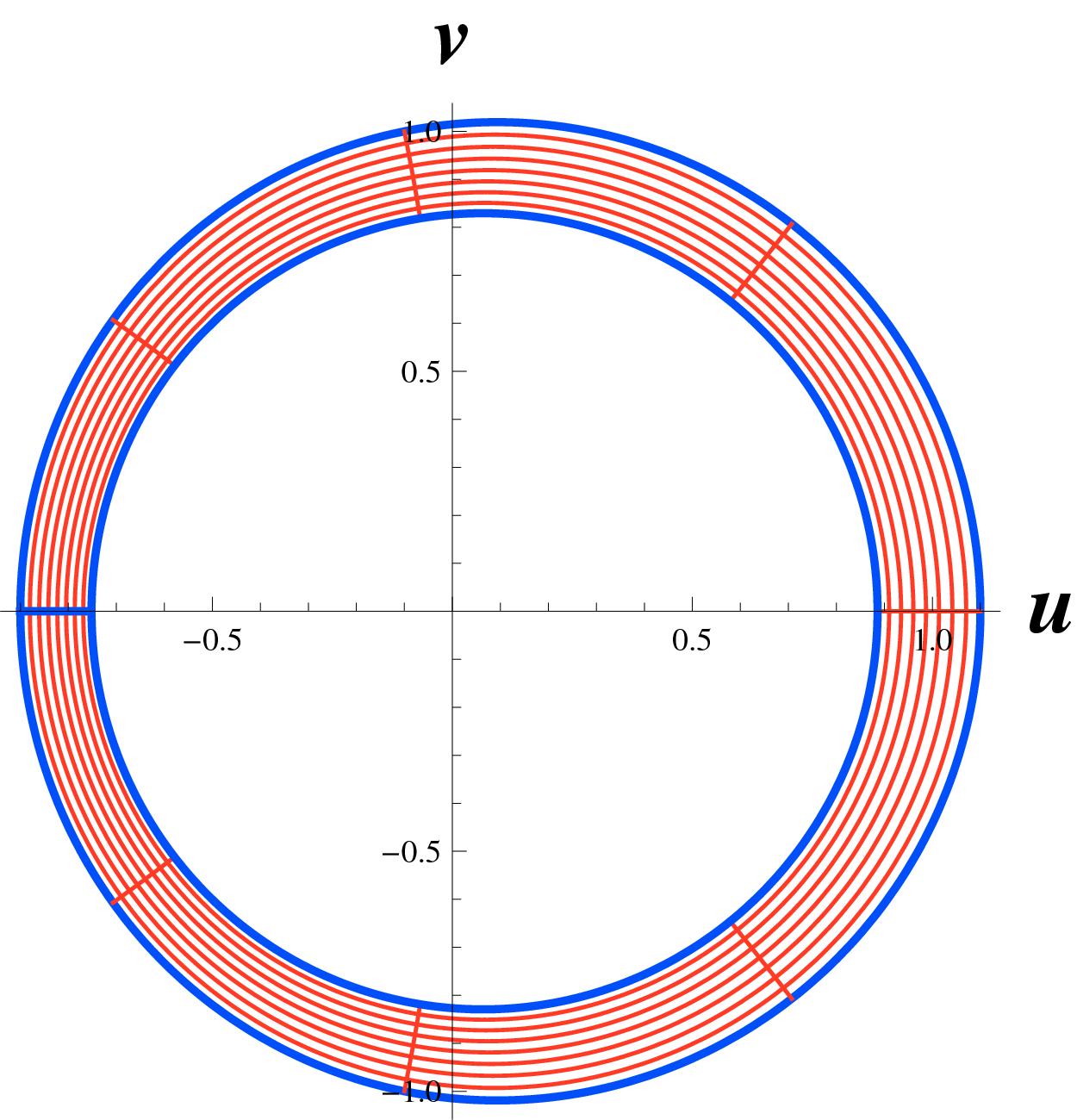}
\caption{(color online) Asymmetric annulus obtained conformally mapping a rectangle of sides $2L_x=1/5$ and $2L_y=2 \pi$ 
centered in the origin. $\alpha=1/10$.}
\label{Fig_cardioid}
\end{center}
\end{figure}

In this case we have:
\beq
\Sigma(x,y) = 4 \alpha ^2 e^{4 x-4 L_x}+4 \alpha  e^{3 x-3 L_x} \cos (y)+e^{2 x-2 L_x}
\eeq

Notice that in this case the width of the ring is not uniform: we call $r_{\mp}$ 
the smallest and largest width of the ring respectively, which read
\beq
r_{\pm} &=&  \pm \alpha  \left(1-e^{-4 {L_x}}\right)-e^{-2 {L_x}} + 1  \ .
\eeq

Their average is insensitive to $\alpha$:
\beq
\bar{r} &\equiv& \frac{r_+ + r_-}{2} = 1-  e^{-2 L_x} \ .
\eeq
Notice also that $1- \bar{r} = e^{-2L_x}=a$  (in this case $a$ is the average inner radius of the
quantum ring).

Using the general results obtained earlier we may write the analytical formula:
\beq
E_{n_x,n_y,s} \approx  \frac{2 \left(\log^2(a)+\pi ^2 {n_x}^2\right) \left({n_y}^2 \log^2(a)+
\pi^2 {n_x}^2\right)}{\pi ^2 \left(a^2-1\right) {n_x}^2 \log (a) + \alpha^2 R_{n_x}(a)}  ,
\label{en_annulus_weyl_2}
\eeq
where 
\beq
R_{n_x}(a) \equiv 2 \pi^2 \left(a^4-1\right) {n_x}^2 \log (a) \left(\log^2(a)+\pi^2 {n_x}^2\right) \nonumber \ .
\eeq
This formula reduces to eq.~(\ref{en_annulus_weyl}) for $\alpha=0$.

\begin{figure}
\begin{center}
\bigskip\bigskip\bigskip
\includegraphics[width=8cm]{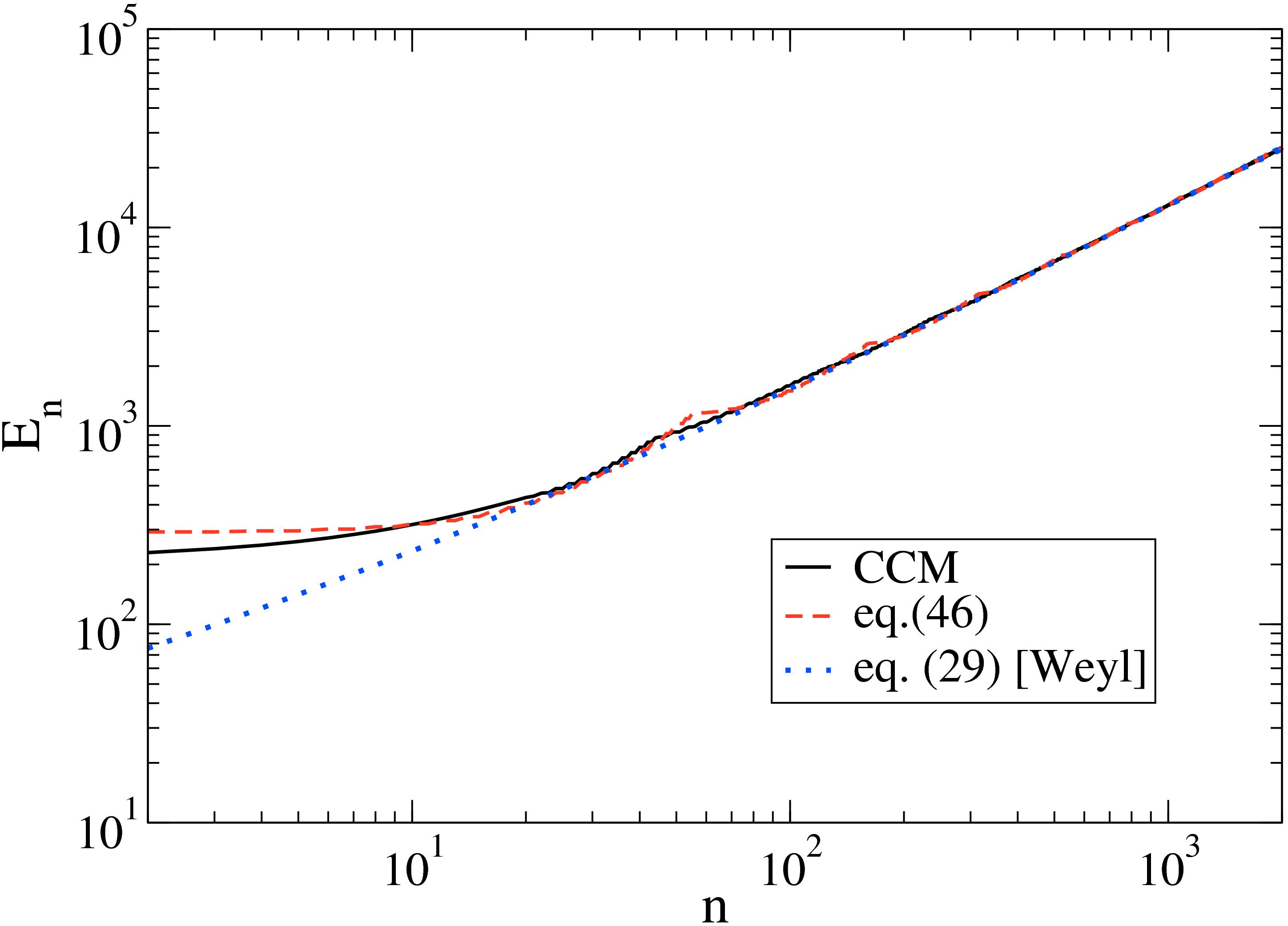}
\caption{(color online)  Energies of the first 2000 states of an annulus obtained conformally mapping a rectangle of 
sides $2L_x=1/5$ and $2L_y=2 \pi$ with the map $f(z) = z + \alpha z^2$ using $\alpha = 1/10$. 
The solild line are the numerical results obtained using CCM; the dashed line corresponds to
the analytical formula, the dotted line is Weyl's law.}
\label{Fig_weyl_robnick}
\end{center}
\end{figure}

In Fig.\ref{Fig_weyl_robnick} we display the energies of the first 2000 states  of an annulus obtained conformally mapping a rectangle of 
sides $2L_x=1/5$ and $2L_y=2 \pi$ with the map $f(z) = z + \alpha z^2$ using $\alpha = 1/10$. The solid line represents
the precise numerical values obtained using CCM with a grid with $N_x=14$ and $N_y=400$; the dashed line corresponds to the results
of the analytical formula of eq.~(\ref{en_annulus_weyl_2}); finally, the dotted line corresponds to Weyl's formula
eq.~(\ref{weyls_conj}). 
In Fig.~\ref{Fig_weyl_robnick} we display  a detail of the previous figure, for highly excited 
states around $n=2000$.

Looking at the figure we see that our analytical formula describes more accurately than Weyl's equation, eq.~(\ref{weyls_conj}),
the low lying part of the spectrum, although the latter describes better the high part of the spectrum. We also notice that
the analytical formula displays a behaviour already observed for the circular annulus: around specific $n$, corresponding to
the opening of a new trasversal mode, the curve displays a small change of slope. While this behaviour is correct for the circular
annulus, where it is indeed observed, neither the numerical results nor (of course) Weyl's expression display such a behaviour.
However we may easily understand the origin of it: the analytical formula eq.~(\ref{en_annulus_weyl_2}) is obtained using 
a resummation of the terms in the Rayleigh-Schr\"odinger series which contains only the expectation values of the conformal density
in a given state. For the circular annulus, this approximation is quite good because there is no mixing between transversal and longitudinal modes
as a result of the rotational symmetry of the domain: this is reflected in the amazing accuracy of the analytical formula in 
this case. For a general annulus, the domain is not invariant under rotation and transverse and longitudinal modes actually mix. 
Already to second order  in perturbation theory, there are terms which allow this mixing, although these are not present in the 
resummed formula.

\begin{figure}
\begin{center}
\bigskip\bigskip\bigskip
\includegraphics[width=8cm]{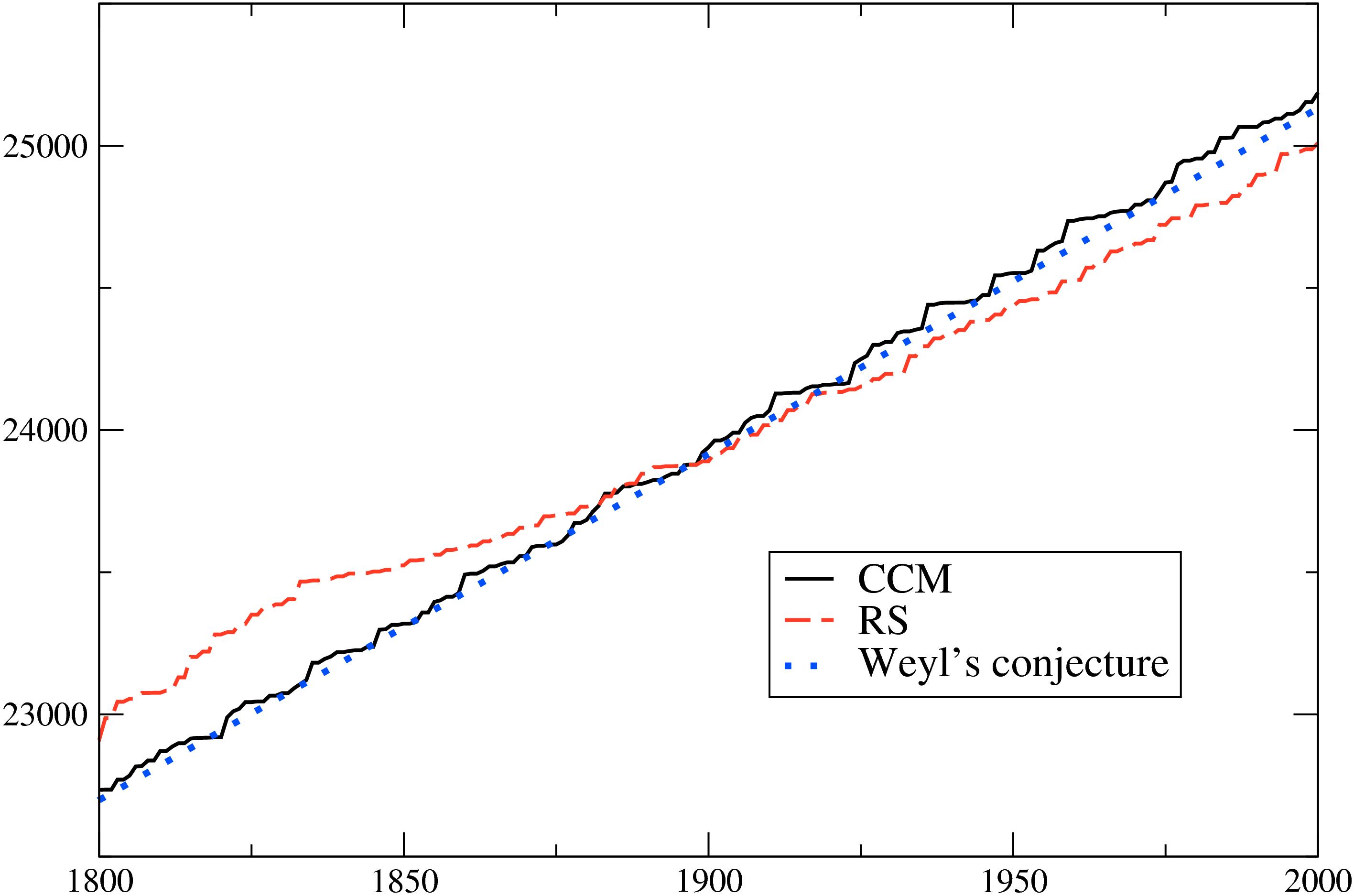}
\caption{(color online)  Detail of the previous figure.}
\label{Fig_weyl_robnick2}
\end{center}
\end{figure}

As a check of the quality of the numerical results obtained with the CCM we have applied a method devised by Berry, ref.~\cite{Berry87}, which
allows one to extract the geometrical feature of a domain from a limited sequence of eigenvalues, using improved eigenvalues sums. 
For the area, for example, one obtains a family of approximating functions whose first few elements read~\cite{Berry87}
\beq
A_0(t) &=& 4\pi t \sum_{n=0}^{N} e^{-E_nt} \\
A_1(t) &=& 4\pi t \sum_{n=0}^{N} e^{-E_nt} (2E_nt-1) \\
A_2(t) &=& 4\pi t \sum_{n=0}^{N} e^{-E_nt} E_nt (2E_nt-3) \\
A_3(t) &=& 4\pi t \sum_{n=0}^{N} e^{-E_nt} E_nt (4E_n^2t^2 - 12 E_nt+3)/3 \\
\dots &=& \dots
\eeq

Analogous expressions are given in Ref.~\cite{Berry87} for the length and constant terms in the asymptotic expasion of 
the function counting the number of states up to a given energy~\cite{BH76}
\beq
\langle n(E) \rangle \approx AE/4\pi - L\sqrt{E}/4\pi + C + \dots 
\eeq
The constant $C$ should vanish for domains with one hole.

\begin{figure}
\begin{center}
\bigskip\bigskip\bigskip
\includegraphics[width=8cm]{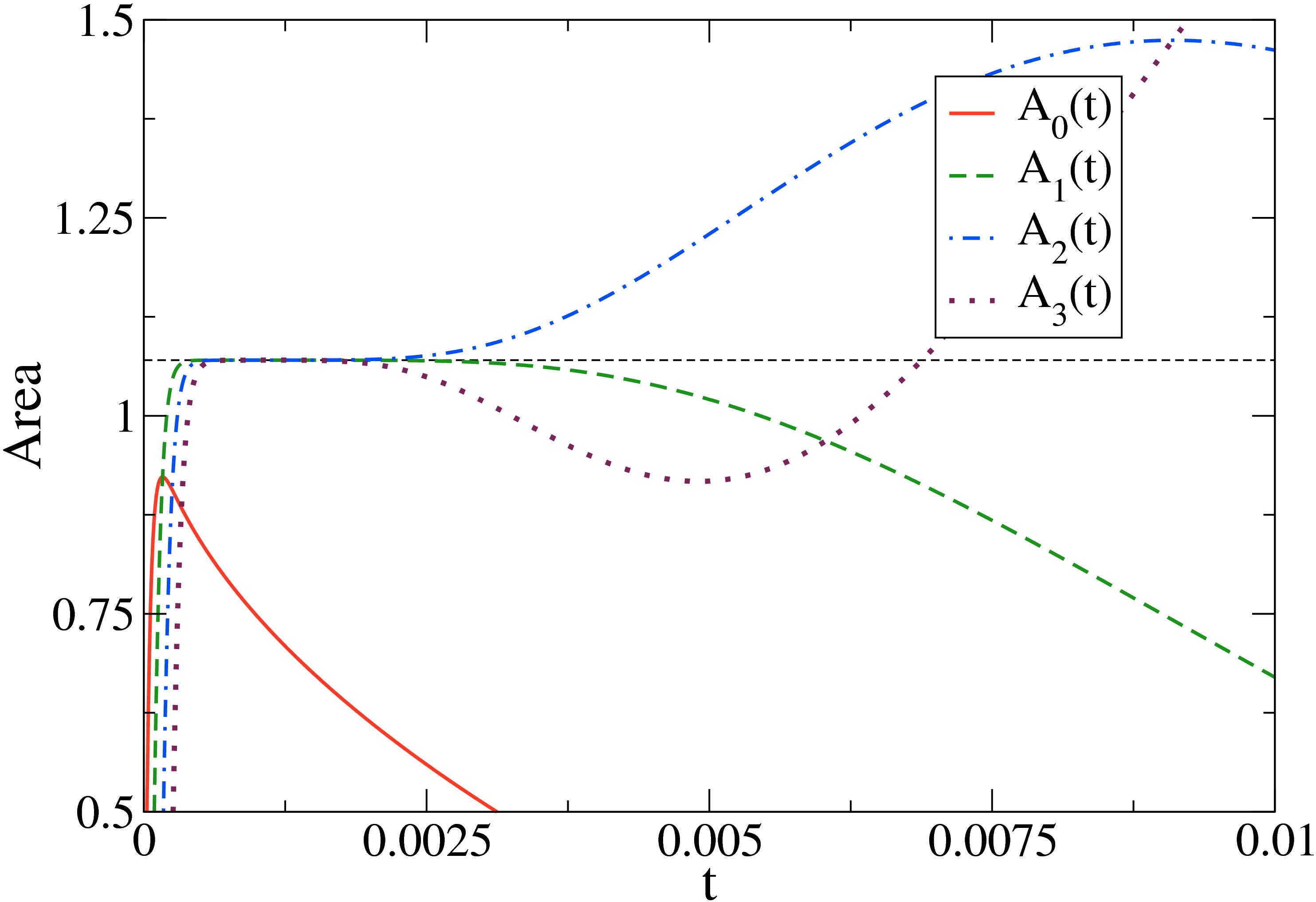}
\caption{(color online) Area approximant functions $A_m(t)$ for the first 2000 eigenvalues obtained with CCM with a grid $14 \times 400$.}
\label{Fig_berryarea}
\end{center}
\end{figure}

\begin{figure}
\begin{center}
\bigskip\bigskip\bigskip
\includegraphics[width=8cm]{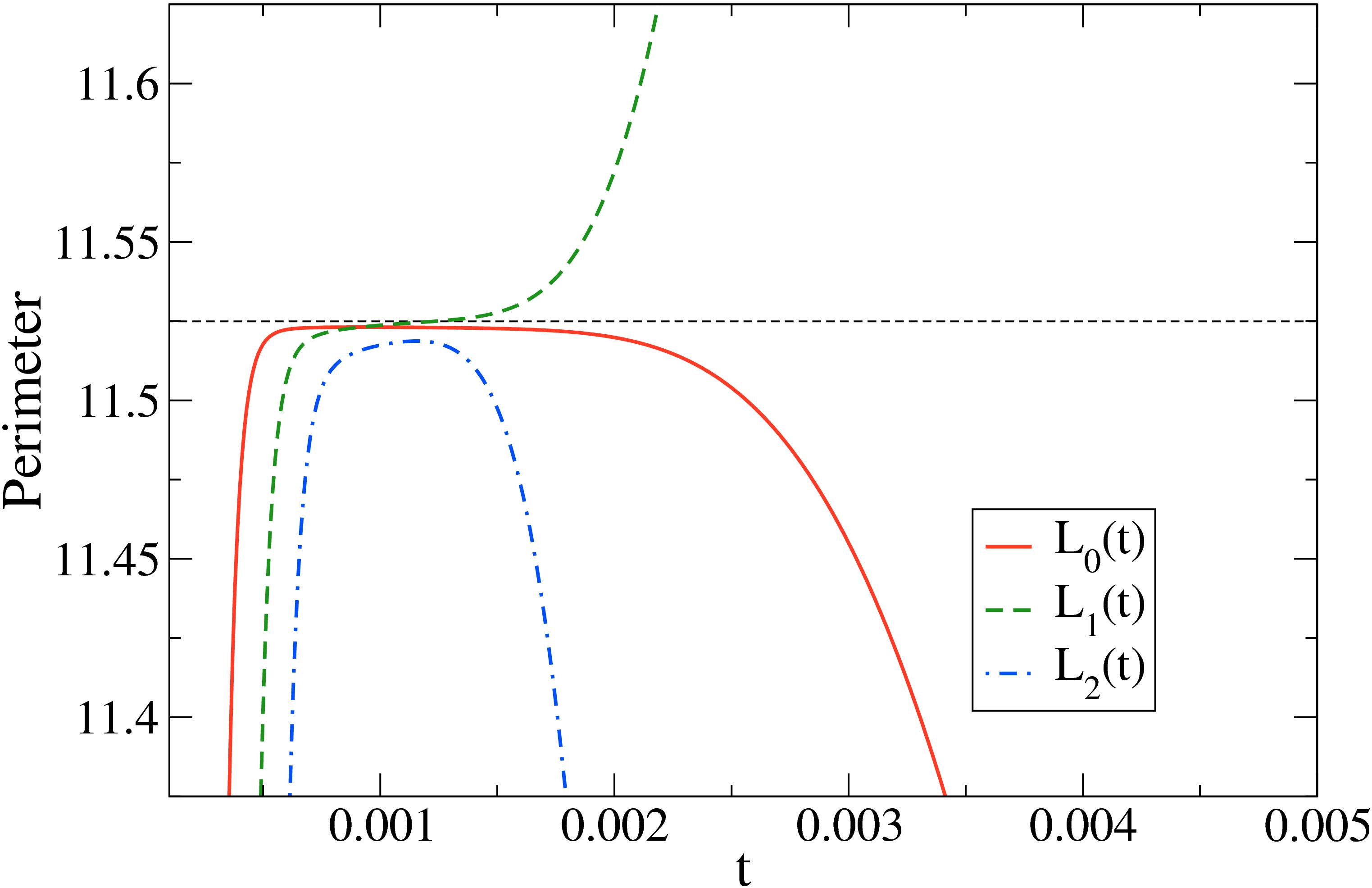}
\caption{(color online) Length approximant functions $L_m(t)$ for the first 2000 eigenvalues obtained with CCM with a grid $14 \times 400$.}
\label{Fig_berryperimeter}
\end{center}
\end{figure}

\begin{figure}
\begin{center}
\bigskip\bigskip\bigskip
\includegraphics[width=8cm]{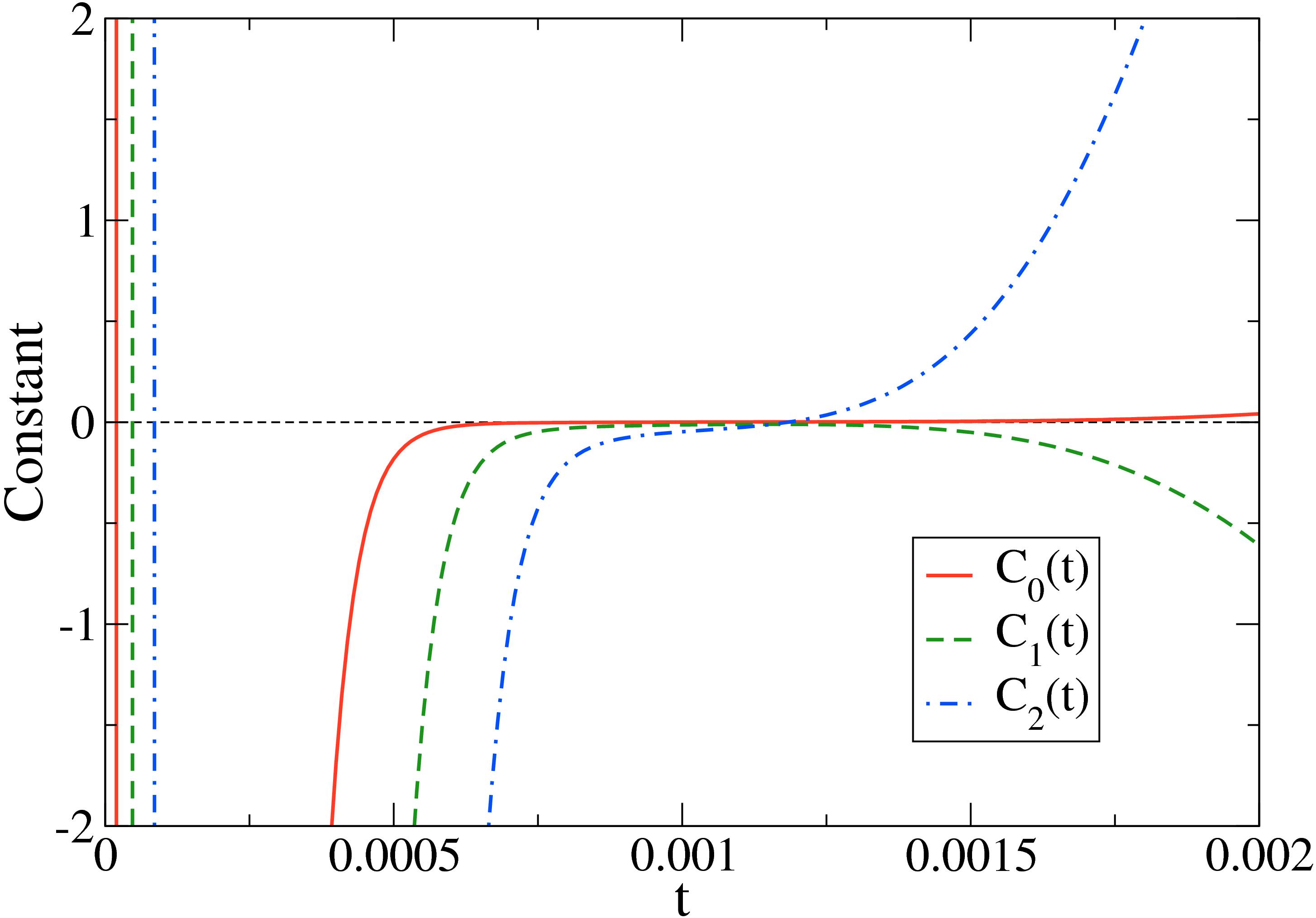}
\caption{(color online) Constant approximant functions $C_m(t)$ for the first 2000 eigenvalues obtained with CCM with a grid $14 \times 400$.}
\label{Fig_berryconstant}
\end{center}
\end{figure}

In figs.\ref{Fig_berryarea},\ref{Fig_berryperimeter} and \ref{Fig_berryconstant} we have displayed the approximants
for the area, perimeter and for the constant term built using the first $2000$ numerical eigenvalues obtained with the
CCM with a grid $14 \times 400$. The horizontal lines are the exact values.

To estimate the optimal value of $t$ where the approximant should be evaluated we minimize the sum of the squares of the
derivatives of each approximant. For example, for the area we minimize the quantity $A_1'(t)^2+A_2'(t)^2+A_3'(t)^2$ (we leave
out $A_0(t)$ because it is clearly less precise). For the length and constant terms we use the approximants displayed in 
the figures.

\begin{table}
\begin{tabular}{|c||c|c|c|}
	\hline
	&  $A$    &   $L$  & $C$   \\
	\hline \hline
method of Ref.~\cite{Berry87}       & 1.07027 & 11.5221  & -0.01738 \\
exact   & 1.07032 & 11.5250  &  0 \\
	\hline
\end{tabular}
\caption{Estimates for the geometrical features of the annulus obtained conformally mapping a rectangle of 
sides $2L_x=1/5$ and $2L_y=2 \pi$ with the map $f(z) = z + \alpha z^2$ using $\alpha = 1/10$ using the method
of Berry,\cite{Berry87} on the numerical results obtained with the CCM. The last row contains the exact results. }
\label{table-2}
\end{table}

As shown in Table~\ref{table-2} the method of Berry allows us to extract the area, perimeter and constant terms of the
annulus with a truly amazing precision. Notice that the exact values of the area and perimeter are obtained
directly using the conformal density:
\beq
A &=&  \int_{\Omega} dxdy \ \Sigma(x,y) \\
L &=& \int_{\partial\Omega} ds \ \sqrt{\Sigma(x,y)} \ ,
\eeq
where $\Omega$ is the rectangle of sides $2L_x=1/5$ and $2L_y=2\pi$ and $\partial \Omega$ is the border of $\Omega$
where Dirichlet bc are obeyed~\footnote{In ref.~\cite{Amore09b} these relations are used to derive Weyl's law from perturbation theory. }.

As we did before, we may also resort to a variational calculation and use the trial state:
\beq
|\chi \rangle &=& c_{0}^{(0)}  \psi_1(x) \chi_0(y) \nonumber \\
&+& \psi_1(x) \sum_{n_y=1}^{N} 
\left[ c_{2 n_y-1}^{(0)}  \phi_{n_y}(y) + c_{2 n_y}^{(0)}  \chi_{n_y}(y)\right]
\eeq
neglecting in first approximation the excited states in the $x$ direction. 
In the case of the circular ring we have seen that only the first term contributes,
due to the rotational invariance of the ground state; this limit corresponds to the
simple formula (\ref{weyl}), which was seen to work remarkably well for thin circular rings.

In the present case we do not expect that the lowest order formula work well even for thin rings,
due to the varying width of the ring, which favors the breaking of the rotational invariance.

We may apply this variational calculation to the Robnik annulus that we have considered before, 
which corresponds to $\alpha = 1/10$ and $L_x = 1/10$, shown in Fig.~\ref{Fig_cardioid}. 
In this case we have
\beq
r_- = 0.148 \ , \ r_+ = 0.214 \ .
\eeq

\begin{figure}
\begin{center}
\bigskip\bigskip\bigskip
\includegraphics[width=8cm]{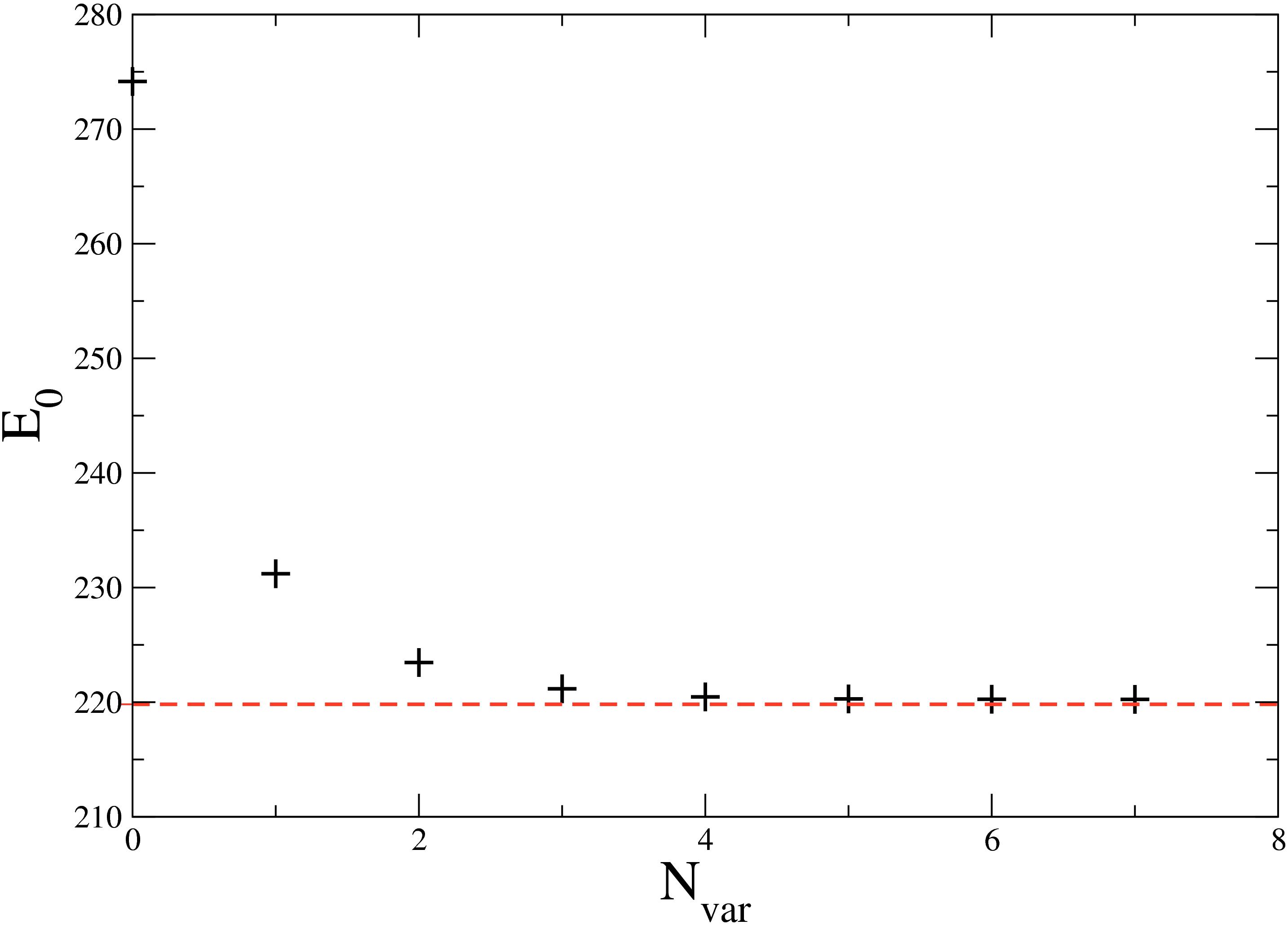}
\caption{(color online) Energy of the ground state of the asymmetric annulus with $L_x=1/10$ and $\alpha=1/10$. The horizontal
line is the precise numerical value obtained with the CCM with a grid with $14 \times 400$; the pluses are the variational
results obtained with a varying number of variational parameters.}
\label{Fig_cardioidring}
\end{center}
\end{figure}

In Fig.~\ref{Fig_cardioidring} we show the variational estimates for the ground state energy of this ring
obtained using a varying number of variational parameters $N_{var}$ (the case $N_{var}=0$ corresponds to the
simple analytical formula (\ref{weyl})); the horizontal line is the precise value obtained with
the Conformal Collocation method with a grid with $N=100$. As anticipated, the eq.~(\ref{weyl}) in this
case provides a poor approximation to the exact energy, leading to a substantial overestimation of it. On the other
hand, the inclusion of few longitudinal modes leads to quite drastic improvement, which can be already appreciated 
for $N_{var}=2$. We should point out that, even letting $N_{var} \Rightarrow \infty$  (assuming that we can perform 
the calculation), we do not expect that the variational energy converge to the exact energy: the reason of this
is that the variational ansatz that we are using includes only longitudinal modes (in the $y$ direction) while completely
neglecting the trasversal ones (in the $x$ direction). For thin rings, the latter are less important, and therefore we are 
able to obtain good (but not arbitrarily good) approximations; for thick rings, on the other hand, only the inclusion
of both modes will lead to acceptable approximations.

\begin{figure}
\begin{center}
\bigskip\bigskip\bigskip
\includegraphics[width=8cm]{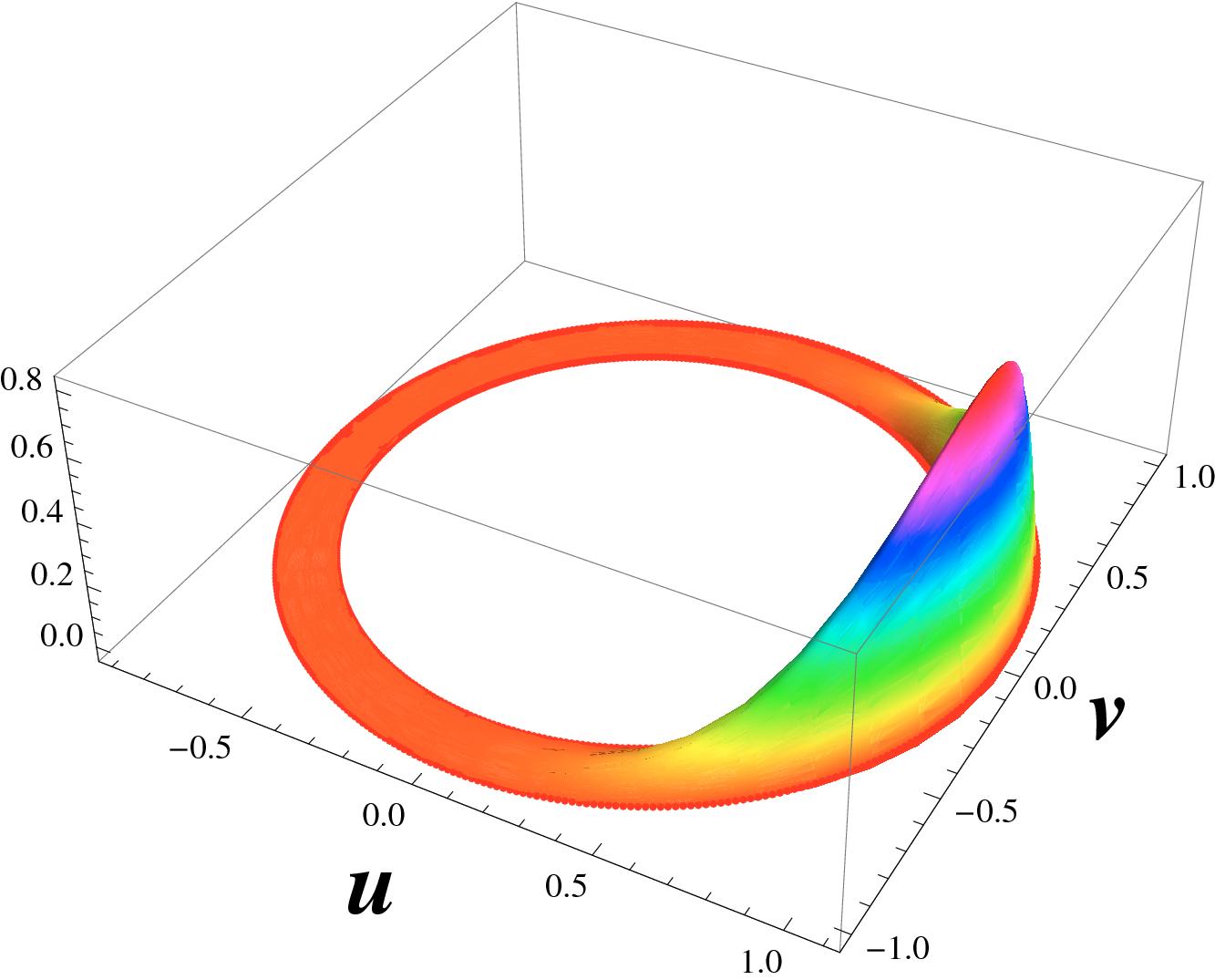}
\caption{(color online) Wave function (multiplied by a factor of $10$) of the  ground state of the Robnik's annulus with $L_x=1/10$ and $\alpha=1/10$.}
\label{Fig_cardioidring3d}
\end{center}
\end{figure}

Finally in Fig. ~\ref{Fig_cardioidring3d} we have plotted the wave function of the ground state of this annulus:
as anticipated, this wave function is not invariant under arbitrary rotations and it is peaked in the region 
where the ring is wider, as expected. The wave function is multiplied by a factor $10$ to better appreciate the localization
in the region where the width of the annulus is larger.

\section{Conclusions}
\label{concl}

In this paper we have extended the powerful techniques of Ref.~\cite{Amore09,Amore09b} to the study of quantum rings;
working on specific examples we have obtained both numerical and analytical results of remarkable precision.
In fact our approach provides a systematic tool for the study of the spectrum of quantum rings of arbitrary shape,
not relying on specific assumptions concerning the shape or the width of the ring. Clearly, the problem of finding the particular
conformal map which sends the rectangle to a specific ring may be attacked numerically, when the explicit expression is
unknown~\cite{Driscoll96,Driscoll05}. 

The results that we obtain in the paper maybe roughly classified into:
\begin{itemize}
\item numerical results obtained using the Conformal Collocation Method (CCM) for a large number of levels ($\approx 2000$); the
accuracy of these results has been verified using Berry's approximants to estimate the geometrical features of the rings with 
a selected number of levels, calculated numerically;
\item variational estimates for the ground state of a quantum ring;
\item explicit analytical formulas for the whole spectrum of a quantum ring; the relation of these formulas with
Weyl's law is also obtained in light of the results of Ref.~\cite{Amore09b};
\end{itemize}

We believe that the importance of our methods thus relies in the possibility of describing with precision both the low and high end 
of the spectrum of a quantum ring and on the possibility of improving this precision in a simple and sistematic fashion. To
the best of our knowledge there are no other approaches in the literature which share these features.

Finally, we wish to mention that the approach that we have described here can be applied directly also to annular drums of variable density
(for simply connected drums this has been done in Ref.~\cite{Amore09b}).

\begin{acknowledgments}
P.Amore ackowledges support of Conacyt through the SNI fellowship.
\end{acknowledgments}


\begin{thebibliography}{}
\bibitem{Amore09} P. Amore, Spectroscopy of drums and quantum billiards: perturbative and non perturbative results,
accepted on the Journal of Mathematical Physics, arXiv:0910.4798v1 [quant-ph] (2009)
\bibitem{Amore08} P. Amore, Journal of Physics {\bf A} 41, 265206 (2008) 
\bibitem{Goldstone92} J.Goldstone and R.L. Jaffe, Phys.Rev.{\bf B} 45, 14100 (1992)
\bibitem{Gridin04} D. Gridin, A. T. I. Adamou, and R. V. Craster, Phys. Rev. {\bf B} 69, 155317 (2004) 
\bibitem{Keller60} J.B. Keller and S.I. Rubinow, Ann. Phys. Leipzig {\bf 9}, 24 (1960)
\bibitem{Keller85} J.B. Keller, SIAM Rev. {\bf 27}, 485 (1985)
\bibitem{Robnik84} M. Robnik, J. Phys.{\bf A} 17, 1049-1074 (1984)
\bibitem{Amore09c} P. Amore, F.M.Fern\'andez, K. Salvo and R.A. Sa\'enz, J. Phys.{\bf A}, 115302 (2009)
\bibitem{Kuttler84} J.R. Kuttler and V.G. Sigillito, SIAM Review {\bf 26} (1984) 163-193
\bibitem{Amore09b} P. Amore, arXiv:0912.1402v1 [math-ph] (2009)
\bibitem{Berry87} M.V.Berry, J. Phys.{\bf A} 20, 2389-2403 (1987)
\bibitem{BH76} Baltes H P and Hilf E R 1976 Spectra of Finite Systems (Mannheim: B-I Wissenschaftsverlag)
\bibitem{Driscoll96} T. A. Driscoll. A MATLAB Toolbox for Schwarz-Christoffel mapping, ACM Trans. Math. Soft. 22 (1996), pp. 168-186.
\bibitem{Driscoll05} T. A. Driscoll. Algorithm 843: Improvements to the Schwarz-Christoffel Toolbox for MATLAB, ACM Trans. Math. Soft. 31 (2005), 239-251.

\end{thebibliography}
\end{document}